\documentclass{elsart}
\usepackage[english]{babel}
\usepackage[latin1]{inputenc}
\usepackage{amsmath}
\usepackage{hhline}
\usepackage{amssymb}
\usepackage{latexsym}
\usepackage{xspace}
\usepackage{fancybox}
\usepackage{natbib}
\usepackage{multicol}
\usepackage{pstricks}
\usepackage{pst-node}
\def\ignore#1{}
\ignore{ \setlength{\oddsidemargin}{0.25in}
\setlength{\textwidth}{6in} \setlength{\topmargin}{-0.25in}
\setlength{\textheight}{8.5in} D}

\def\calS0{{\cal S}_0}

\def\cald{\bf D}
\def\calq{\bf Q}

\def\wnnw#1{\omega_{#1}}

\def\sat#1{{\it Sat}({#1})}
\def\entails#1#2{{#1}\vdash{#2}}
\def\contains#1#2#3{{#2}\subseteq_{#1}{#3}}

\newtheorem{definition}{Definition}
\newtheorem{theorem}{Theorem}
\newtheorem{corollary}{Corollary}
\newtheorem{proposition}{Proposition}

\newtheorem{example}{Example}

\newenvironment{proof}{\noindent{\em Proof.}}{\hfill\eop}
\newcommand{\eop}{\hbox{\hskip6pt\vrule height 6pt width 6pt}}

\begin{document}
\begin{frontmatter}
\title{\bf Semantic Optimization Techniques for Preference Queries\thanksref{NSF}}
\thanks[NSF]{Research supported by
NSF Grant IIS-0307434. An earlier version of some of the results in this paper was presented
in \cite{ChCDB04}.}

\author{Jan Chomicki}
\address{Dept. of Computer Science and Engineering, University
at Buffalo,Buffalo, NY 14260-2000, {\tt
chomicki@cse.buffalo.edu}}


\begin{abstract}
Preference queries are relational algebra or SQL queries that
contain occurrences of the winnow operator ({\em find the most
preferred tuples in a given relation}).
Such queries are parameterized by specific {\em preference relations}.
Semantic optimization techniques make use of integrity constraints
holding in the database.
In the context of semantic optimization of preference queries, we identify two fundamental properties:
{\em containment} of preference relations relative to integrity
constraints and {\em satisfaction of order axioms} relative to
integrity constraints.
We show numerous applications of those notions to
preference query evaluation and optimization.
As integrity constraints, we consider {\em constraint-generating
dependencies}, a class generalizing functional dependencies.
We demonstrate that the problems of containment and satisfaction
of order axioms can be captured as specific instances of
constraint-generating dependency entailment.
This makes it possible to formulate necessary and sufficient conditions for the applicability of our techniques
as {\em constraint validity} problems.
We characterize the computational complexity of such problems.
\end{abstract}
\end{frontmatter}
\section{Introduction}\label{sec: introduction}
The notion of {\em preference} is becoming more and more
ubiquitous in present-day information systems. Preferences are
primarily used to filter and personalize the information reaching
the users of such systems. In database systems, preferences are
usually captured as {\em preference relations} that are used to
build {\em preference queries}
\cite{ChEDBT02,ChTODS03,Kie02,KiKo02}. From a formal point of
view, preference relations are simply binary relations defined on
query answers. Such relations provide an abstract, generic way to
talk about a variety of concepts like priority, importance,
relevance, timeliness, reliability etc. Preference relations can
be defined using logical formulas \cite{ChEDBT02,ChTODS03} or
special preference constructors \cite{Kie02} (preference
constructors can be expressed using logical formulas). The
embedding of preference relations into relational query languages
is typically provided through a relational operator that selects
from its argument relation the set of the {\em most preferred
tuples}, according to a given preference relation. This operator
has been variously called {\em winnow} (the term we use here)
\cite{ChEDBT02,ChTODS03}, BMO \cite{Kie02}, and Best
\cite{ToCi02}. It is also implicit in {\em skyline queries}
\cite{BoKoSt01}. Being a relational operator, winnow can clearly
be combined with other relational operators, in order to express
complex preference queries.

\begin{example}\label{ex:book}
We introduce an example used throughout the paper.
Consider the relation $Book(ISBN,Vendor,Price)$
and the following preference relation $\succ_{C_1}$ between {\em Book} tuples:
\begin{quote}
{\em prefer one Book tuple to another if and only if their ISBNs are the same and the Price of the
first is lower.}
\ignore{
{\em if two tuples have the same ISBN and different Price, prefer the one with the lower Price}.}
\end{quote}
Consider the instance $r_1$ of $Book$ in Figure \ref{fig:book}.
Then the winnow operator $\wnnw{C_1}$ returns the set of tuples in
Figure \ref{fig:winnow}.

\begin{figure}[htb]
\centering
\begin{tabular}{|l|l|l|}
\hline
{\em ISBN} &{\em Vendor} &{\em Price}\\\hline
0679726691 & BooksForLess & \$14.75\\
0679726691 & LowestPrices & \$13.50\\
0679726691 & QualityBooks & \$18.80\\
0062059041 & BooksForLess & \$7.30\\
0374164770 & LowestPrices & \$21.88\\\hline
\end{tabular}
\caption{The Book relation}
\label{fig:book}
\end{figure}

\begin{figure}[htb]
\centering
\begin{tabular}{|l|l|l|}
\hline
{\em ISBN} &{\em Vendor} &{\em Price}\\\hline
0679726691 & LowestPrices & \$13.50\\
0062059041 & BooksForLess & \$7.30\\
0374164770 & LowestPrices & \$21.88\\\hline
\end{tabular}
\caption{The result of winnow}
\label{fig:winnow}
\end{figure}
\end{example}
\begin{example}
The above example is a one-dimensional skyline query.
To see an example of a two-dimensional skyline, consider
the schema of {\em Book\/} expanded by another attribute {\em Rating\/}.
Define the following preference relation $\succ_{C_2}$:
\begin{quote}
{\em prefer one Book tuple to another if and only if their ISBNs are the same and the Price of the
first is lower and the Rating of the first is not lower, or the Price of the first is not higher
and the Rating of the first is higher.}
\end{quote}
Then $\wnnw{C_2}$ is equivalent to the following skyline (in the terminology of \cite{BoKoSt01}):
\begin{verbatim}
    SKYLINE ISBN DIFF, Price MIN, Rating MAX.
\end{verbatim}
The above notation indicates that only books with the same ISBN
should be compared, that Price should be minimized, and Rating
maximized. In fact, the tuples in the skyline satisfy the property
of {\em Pareto-optimality}, well known in economics.
\end{example}

Preference queries can be reformulated in relational algebra or SQL, and thus
optimized and evaluated using standard relational techniques. However, it has
been recognized that specialized evaluation and optimization techniques promise
in this context performance improvements that are otherwise unavailable.
A number of new algorithms for the evaluation of skyline queries (a special class
of preference queries) have been proposed \cite{BalEDBT04,BoKoSt01,ChGoGrLi03,GoShGr05,KoRaRo02,PaTaFuSe03}.
Some of them can be used to evaluate more general preference queries \cite{BalVLDB04,ChTODS03}.
Also, algebraic laws that characterize the interaction of winnow with the
standard operators of relational algebra have been formulated \cite{ChTODS03,KiHa02,KiHa03}.
Such laws provide a foundation for the rewriting of preference queries.
For instance, necessary and sufficient conditions for pushing a selection through
winnow are described in \cite{ChTODS03}.
The algebraic laws cannot be applied unconditionally.
In fact, the preconditions of their applications refer to the {\em validity}
of certain {\em constraint formulas}.

In this paper, we pursue the line of research from \cite{ChTODS03}
a bit further. We study {\em semantic optimization} of preference
queries. Semantic query optimization has been extensively studied
for relational and deductive databases \cite{CHGrMi90}. As a
result, a body of techniques dealing with specific query
transformations like join elimination and introduction, predicate
introduction etc. has been developed. We view semantic query
optimization very broadly and classify as {\em semantic} any query
optimization technique that makes use of integrity constraints. 
In the context of semantic optimization of preference queries,
we identify two fundamental {\em semantic properties}:
{\em containment} of preference relations relative to  integrity
constraints and {\em satisfaction of order axioms} relative to
integrity constraints.
We show that those notions make it possible to formulate 
semantic query optimization techniques for preference queries
in a uniform way.

We focus on  the winnow
operator. Despite the presence of specialized evaluation
techniques, winnow, being essentially an anti-join,  is still quite an expensive operation. We
develop optimizing techniques that:
\begin{enumerate}
\item remove redundant occurrences of winnow;
\item coalesce consecutive applications of winnow;
\item recognize when more efficient evaluation of winnow is possible.
\end{enumerate}
More efficient evaluation of winnow can be achieved, for example,
if the given preference relation is a {\em weak order} (a
negatively transitive strict partial order). We show that even
when the preference relation is not a weak order (as in Example
\ref{ex:book}), it may become  a weak order on the
relations satisfying certain integrity constraints. We show a very
simple, single-pass algorithm for evaluating winnow under those
conditions. We also pay attention to the issue of satisfaction of
integrity constraints in the result of applying winnow. In fact,
some integrity constraints may hold in the result of winnow, even though
they do not hold in the relation to which winnow is applied.
Combined with known results about the preservation of integrity
constraints by relational algebra operators \cite{Klu80,KlPr82},
our results provide a way for optimizing not only single
occurrences of winnow but also complex preference queries. 

As integrity constraints, we consider {\em constraint-generating
dependencies} \cite{BaChWo99}, a class generalizing functional dependencies.
Constraint-generating dependencies seem particularly well matched with preference queries,
since both the former and the latter are formulated using constraints.
We demonstrate that the problems of containment of preference relations and satisfaction
of order axioms, relative to integrity constraints, can be captured as specific instances of
dependency entailment.
Our approach makes it possible to formulate necessary and sufficient conditions for the applicability of the
proposed semantic query optimization techniques
as {\em constraint validity} problems and precisely characterize the computational complexity of such problems,
partly adopting some of the results of \cite{BaChWo99}.

The plan of the paper is as follows. In Section \ref{sec:basic}, we
provide background material on preference queries and constraint-generating
dependencies.
In Section \ref{sec:relative}, we introduce two basic semantic properties:
relative containment and relative satisfaction of order axioms.
In Section \ref{sec:redundant}, we address the issue of eliminating redundant
occurrences of winnow. In Section \ref{sec:weak}, we study weak
orders. In Section \ref{sec:propagation}, we characterize
dependencies holding in the result of winnow. 
In Section \ref{sec:complexity}, we consider the computational complexity of
the semantic properties  studied in the present paper.
We discuss related work in Section \ref{sec:related}, and conclude in
Section \ref{sec:concl}.

\section{Basic notions}\label{sec:basic}

We are working in the context of the relational model of data.
For concreteness, we consider two infinite domains: $\cald$ (uninterpreted constants) and $\calq$ (rational numbers).
Other domains could be considered as well without influencing most of the results of the paper.
We assume that database instances are finite.
Additionally,
we have the standard built-in predicates.
We refer to relation attributes using their names or positions.

We define {\em constraints} to be quantifier-free formulas over some signature
of built-in operators, interpreted over a fixed domain - in our case
$\cald$ or $\calq$. 
We will allow both atomic- and tuple-valued variables in constraints.
The notation $t[X]$ denotes the fragment of a tuple $t$ consisting
of the values of the attributes in the set $X$.

\subsection{Preference relations}
\begin{definition}\label{def:prefrel}
Given a relation schema $R(A_1 \cdots A_k)$
such that $U_i$, $1\leq i\leq k$, is the domain (either $\cald$ or $\calq$)
of the attribute $A_i$, a relation $\succ$ is a {\em preference relation over $R$}
if it is a subset of $(U_1\times\cdots\times U_k)\times (U_1\times\cdots\times U_k)$.
\end{definition}

Intuitively, $\succ$ will be a binary relation between tuples from the
same (database) relation.
We say that a tuple $t_1$ {\em dominates} a tuple $t_2$
in $\succ$ if $t_1\succ t_2$.

Typical properties of the relation $\succ$ include:
\begin{itemize}
\item {\em irreflexivity}: $\forall x.\ x\not\succ x,$
\item {\em asymmetry}: $\forall x,y.\ x\succ y\Rightarrow y\not\succ x,$
\item {\em transitivity}: $\forall x,y,z.\ (x\succ y \wedge y\succ z)\Rightarrow x\succ z,$
\item {\em negative transitivity}: $\forall x,y,z.\ (x\not\succ y \wedge y\not\succ z)\Rightarrow x\not\succ z,$
\item {\em connectivity}: $\forall x,y.\ x\succ y\vee y\succ x \vee x=y.$
\end{itemize}

The relation $\succ$ is:
\begin{itemize}
\item a {\em strict partial order} if it is
irreflexive and transitive (thus also asymmetric);
\item a {\em weak order} if it is a
negatively transitive strict partial order;
\item a {\em total order} if it is
a connected strict partial order.
\end{itemize}

At this point, we do not assume any properties
of $\succ$, although in most applications it will satisfy at least the properties of {\em strict partial
order}.

\begin{definition}\label{def:prefformula}
A {\em preference formula (pf)} $C(t_1,t_2)$ is a first-order
formula defining a preference relation $\succ_C$ in the standard
sense, namely
\[t_1\succ_C t_2\;{\rm iff}\; C(t_1,t_2).\]
An {\em intrinsic preference formula (ipf)} is a preference formula that
uses only built-in predicates.
\end{definition}

We will limit our attention to preference relations defined using intrinsic preference
formulas. Most preference relations of this form. Moreover,
for intrinsic preference relations testing a pair of tuples for dominance can be entirely done
on the basis of the contents of those tuples; no database queries need to be evaluated.

Because we consider two specific domains, $\cald$ and $\calq$, we will have two
kinds of variables, $\cald$-variables and $\calq$-variables, and two kinds
of atomic formulas:
\begin{itemize}
\item {\em equality  constraints}: $x=y$, $x\not=y$, $x=c$, or $x\not= c$, where
$x$ and $y$ are $\cald$-variables, and $c$ is an uninterpreted constant;
\item {\em rational-order constraints}: $x\theta y$ or $x\theta c$, where
$\theta\in\{=,\not=,<,>,\leq,\geq\}$, $x$ and $y$ are $\calq$-variables, and $c$ is a
rational number.
\end{itemize}

Without loss of generality, we
will assume that ipfs are in DNF (Disjunctive Normal Form) and
quantifier-free (the theories involving the above domains admit
quantifier elimination).
We also assume that atomic formulas are closed under negation (also
satisfied by the above theories).
An ipf whose all atomic formulas are equality  (resp. rational-order)
constraints will be called an {\em equality} (resp. {\em rational-order}) ipf.
If both equality and rational-order constraints are allowed in a formula, the
formula will be called {\em equality/rational-order}.
Clearly, ipfs are a special case of general constraints
\cite{CDB00}, and define {\em fixed}, although possibly infinite,
relations.
By using the notation $\succ_C$ for a preference relation, we assume that there is
an underlying preference formula $C$.

Every preference relation $\succ_C$ generates an indifference relation $\sim_C$:
two tuples $t_1$ and $t_2$ are {\em indifferent}
($t_1\sim_C t_2$) if
neither is preferred to the other one, i.e.,
$t_1\not\succ_C t_2$ and $t_2\not\succ_C t_1$.

\begin{proposition}\label{prop:indiff}
For every preference relation $\succ_C$, every relation $r$ and every tuple $t_1,t_2\in\wnnw{C}(r)$,
we have $t_1=t_2$ or $t_1\sim_C t_2$.
\end{proposition}

Complex preference relations can be easily defined using Boolean connectives.
Here we define a special operator: prioritized composition.
The prioritized composition $\succ_{C_1} \rhd \succ_{C_2}$ has the following
intuitive reading: {\em prefer according to $\succ_{C_2}$ unless $\succ_{C_1}$
is applicable.}

\begin{definition}\label{def:priority}
Consider two preference relations $\succ_{C_1}$ and $\succ_{C_2}$
defined over the same schema $R$.
The {\em prioritized composition} $\succ_{C_{1,2}}\ =\ \succ_{C_1} \rhd \succ_{C_2}$ of 
$\succ_{C_1}$ and $\succ_{C_2}$ is a preference relation over $R$ defined as:
\[t_1\succ_{C_{1,2}}t_2\equiv t_1\succ_{C_1}t_2 \vee
(t_1\sim_{C_1}t_2\wedge t_1\succ_{C_2}t_2).\]
\end{definition}

\subsection{Winnow}
We define now an algebraic operator that picks from a given relation the
set of the {\em most preferred tuples}, according to a given preference formula.
\begin{definition}\label{def:winnow}
If $R$ is a relation schema and $C$ a preference
formula defining a preference relation $\succ_C$ over $R$,
then the {\em winnow operator} is written as $\wnnw{C}(R)$,
and for every instance $r$ of $R$:
\[\wnnw{C}(r)=\{t\in r\mid\neg \exists t'\in r.\ t'\succ_C t\}.\]
\end{definition}

A preference query is a relational algebra query containing at least
one occurrence of the winnow operator.

\begin{example}\label{ex:book:02}
Consider the relation $Book(ISBN,Vendor,Price)$ (Example
\ref{ex:book}). The preference relation $\succ_{C_1}$ from this
example can be defined using the rational-order ipf $C_1$:
\[(i,v,p)\succ_{C_1}(i',v',p') \equiv i=i' \wedge p<p'.\]
The answer to the preference query
$\wnnw{C_1}(Book)$
provides for every book the information about the vendors offering the
lowest price for that book.
Note that the preference relation $\succ_{C_1}$ is a strict partial order.
\end{example}

\begin{example}
To see another kind of preferences, consider the following preference relation $\succ_{C_3}$:
\begin{quote}
I prefer Warsaw to any other city and prefer any city to Moscow.
\end{quote}
This preference relation can be formulated as an equality ipf $C_3$:
\[x\succ_{C_3} y\equiv x={\rm 'Warsaw'} \wedge y\not={\rm 'Warsaw'}\vee
x\not={\rm 'Moscow'}\wedge y={\rm 'Moscow'}.\]
\end{example}

\subsection{Constraint-generating dependencies}
We assume that we are working in the context of a single relation
schema $R$ and all the integrity constraints are over that schema. The
set of all instances of $R$ satisfying a set of integrity
constraints $F$ is denoted as $\sat{F}$. We say that $F$ {\em
entails} an integrity constraint $f$, written $\entails{F}{f}$, if every instance satisfying
$F$ also satisfies $f$. 

Remember that constraints are arbitrary quantifier-free formulas over some constraint
theory - here ${\cald}$ or ${\calq}$.

\begin{definition}\label{def:cgd}{\rm \cite{BaChWo99}}
A {\em constraint-generating dependency (CGD)} can be expressed a formula of the following form:
\[\forall t_1.\ldots\forall t_k.\ R(t_1)\wedge\cdots\wedge R(t_k)\wedge \gamma(t_1,\ldots t_k)
\Rightarrow \gamma'(t_1,\ldots t_k)\] where $\gamma(t_1,\ldots
t_k)$ and $\gamma'(t_1,\ldots t_k)$ are constraints. Such a dependency is called a $k$-dependency.
\end{definition}
CGDs are equivalent to denial constraints.
Functional dependencies (FDs) are $2$-CGDs,
because a functional dependency (FD) $f\equiv X\rightarrow Y$, where $X$ and $Y$ are sets
of attributes of $R$, can be written down as the following logic formula:
\[\forall t_1.\forall t_2.\ R(t_1)\wedge R(t_2)\wedge t_1[X]=t_2[X] \Rightarrow
t_1[Y]=t_2[Y].\]
Note that the set of attributes $X$ in $X\rightarrow Y$ may be empty, meaning
that each attribute in $Y$ can assume only a single value.

\begin{example}
We give here further examples of CGDs. Consider the relation {\it
Emp} with attributes {\it Name}, {\it Salary}, and {\it Manager},
with {\it Name} being the primary key. The constraint that {\it no
employee can have a salary greater that that of her manager} is a
CGD:
\[\forall n,s,m,s',m'.~{\it Emp\/}(n,s,m)\wedge{\it Emp\/}(m,s',m')\Rightarrow
s\leq s'.\]
Similarly, single-tuple constraints ({\tt CHECK} constraints in SQL2) are
a special case of CGDs. For example, the constraint that
{\em no employee can have a salary over \$200000} is expressed as:
\[\forall n,s,m.~{\it Emp\/}(n,s,m)\Rightarrow s\leq 200000.\]
\end{example}

The paper \cite{BaChWo99}
contains an effective reduction using {\em symmetrization} from
the entailment of CGDs to the validity of  $\forall$-formulas (or, equivalently,
to the unsatisfiability of quantifier-free formulas) in the
underlying constraint theory.
This reduction is descibed in Section \ref{sec:complexity}.
 A similar construction using {\em
symbol mappings} is presented in \cite{ZhOz97}.

\section{Properties relative to integrity constraints}\label{sec:relative}

We define here two properties fundamental to semantic query optimization
of preference queries: containment of preference relations and satisfaction
of order axioms.

\begin{definition}\label{def:containment}
A preference relation $\succ_{C_1}$ over a schema $R$ is {\em contained} in a preference relation
$\succ_{C_2}$ over the same schema {\em relative to a set of integrity constraints $F$}, written
as $\contains{F}{\succ_{C_1}}{\succ_{C_2}}$ if 
\[\forall r\in \sat{F}. \ \forall t_1,t_2\in r.\ t_1\succ_{C_1} t_2\Rightarrow t_1\succ_{C_2} t_2.\]
\end{definition}
Clearly, $\contains{F}{\succ_{C_1}}{\succ_{C_2}}$ iff $\entails{F}{d_0^{C_1,C_2}}$, where

\begin{center}
\fbox{$d_0^{C_1,C_2}:\ \forall t_1,t_2.\ R(t_1)\wedge R(t_2)\wedge t_1\succ_{C_1} t_2\Rightarrow t_1\succ_{C_2} t_2.$}
\end{center}

Satisfaction of order axioms relative to integrity constraints is defined similarly --
by relativizing the universal quantifiers in the axioms.
Since in this paper we are interested in strict partial and weak orders, we define the following.

\begin{definition}\label{def:SPO}
A preference $\succ_{C}$ over a schema $R$ is a {\em strict partial order relative to a set of integrity constraints $F$}
if:
\[\begin{array}{l}
\forall r\in \sat{F}. \ \forall t\in r.\ t\not\succ_C t\\
\forall r\in \sat{F}. \ \forall t_1,t_2,t_3\in r.\ t_1\succ_C t_2\wedge t_2\succ_C t_3\Rightarrow t_1\succ_C t_3.\\
\end{array}\]
\end{definition}

\begin{definition}\label{def:WO}
A preference $\succ_{C}$ over a schema $R$ is a {\em weak order relative to a set of integrity constraints $F$}
if it is a strict partial order relative to $F$ and 
\[\begin{array}{l}
\forall r\in \sat{F}. \ \forall t_1,t_2,t_3\in r.\ t_1\not\succ_C t_2\wedge t_2\not\succ_C t_3
\Rightarrow t_1\not\succ_C t_3.\\
\end{array}\]
\end{definition}

Again, it is clear that the above properties can be expressed in terms of the entailment of CGDs.

\section{Eliminating redundant occurrences of winnow}\label{sec:redundant}

We consider here two situations in which an occurrence of winnow in a preference query may be eliminated.
The first case is that when a single application of winnow does not remove any tuples, and is thus redundant.
The second case is more subtle: the interaction between two consecutive applications
of winnow is such that one can be eliminated.

Given an instance $r$ of $R$, the operator $\wnnw{C}$ is redundant if $\wnnw{C}(r)=r$.
If we consider the class of all instances of $R$, then such an operator is redundant
for every instance iff $\succ_C$ is an empty relation.
The latter holds iff $C$ is unsatisfiable.
However, we are interested only in the instances satisfying a given set of integrity
constraints.

\begin{definition}
Given a set of integrity constraints $F$, the operator $\wnnw{C}$ is {\em redundant
relative to a set of integrity constraints $F$} if $\forall r\in\sat{F}$, $\wnnw{C}(r)=r$.
\end{definition}

\begin{theorem}\label{th:redundant}
$\wnnw{C}$ is redundant relative to a set of FDs $F$ iff $\contains{F}{\succ_C}{\succ_{False}}$
where 
\[t_1\succ_{False} t_2 \equiv False.\]
\end{theorem}
\begin{proof}
Assume $t_1,t_2\in r$ for some $r\in\sat{F}$ and $t_1\succ_C t_2$.
Then  $t_2\not\in\wnnw{C}(r)$.
In the other direction, assume that for some $r\in\sat{F}$, $\wnnw{C}(r)\subset r$.
Thus, there must be $t_1,t_2\in r$ such that $t_1\succ_C t_2$.
\end{proof}

Clearly, $\wnnw{C}$ is redundant relative to $F$ iff $F$ entails the following CGD:

\begin{center}
\fbox{$d_1^C:\ \forall t_1,t_2.\ R(t_1)\wedge R(t_2)\Rightarrow t_1\not\succ_{C} t_2.$}
\end{center}
The CGD $d_1^C$ always holds in the result of winnow $\wnnw{C}(R)$ and simply says
that all the tuples in this result are mutually indifferent.

\begin{example}\label{ex:book:03}
Consider Example \ref{ex:book:02} in which the FD $ISBN\rightarrow Price$ holds. 
$\wnnw{C_1}$ is redundant relative to $ISBN\rightarrow Price$ because
this dependency entails (is, in fact, equivalent to) the dependency
\[\forall i_1,v_1,p_1,i_2,v_2,p_2.\ Book(i_1,v_1,p_1)\wedge Book(i_2,v_2,p_2)\Rightarrow
i_1\not= i_2 \vee p_1\geq p_2.\]
\end{example}

The second case where an occurrence of winnow can be eliminated is as follows.

\begin{theorem}\label{th:nested}
Assume $F$ is a set of integrity constraints over a schema $R$.
If $\succ_{C_1}$ and $\succ_{C_2}$ are preference relations over $R$ such that 
$\succ_{C_1}$ and $\succ_{C_2}$ are strict partial orders relative to $F$ and 
$\contains{F}{\succ_{C_1}}{\succ_{C_2}}$,
then for all instances $r\in \sat{F}$:
\[\wnnw{C_1}(\wnnw{C_2}(r))=\wnnw{C_2}(\wnnw{C_1}(r))=\wnnw{C_2}(r).\]
\end{theorem}
\begin{proof}
Theorem 6.1 in \cite{ChTODS03} is a similar result which does not, however,
relativize the properties of the given preference relations to the set of instances
satisfying the given integrity constraints.
The proof of that result can be easily adapted here.
\end{proof}

Note that in the case of strict partial orders, Theorem \ref{th:nested} implies one
direction of Theorem \ref{th:redundant}.

\section{Weak orders}\label{sec:weak}

We have defined weak orders as negatively transitive strict partial orders.
Equivalently, they can be defined as strict partial orders for which the
indifference relation is transitive.
Intuitively, a weak order consists of a number (perhaps infinite) of linearly ordered layers.
In each layer, all the elements are mutually indifferent and they are all above all the elements in lower
layers.
\begin{example}\label{ex:book:04}
In the preference relation $\succ_{C_1}$ in Example \ref{ex:book:02}, the first, second and third tuples are
indifferent with the fourth and fifth tuples. However, the first tuple is preferred to the
second, violating the transitivity of indifference. Therefore, the preference relation $\succ_{C_1}$
is not a weak order.
\end{example}
\begin{example}
A preference relation $\succ_{C_f}$, defined as
\[ x\succ_{C_f} y \equiv f(x)>f(y)\]
for some real-valued function $f$, is a weak order but not necessarily a total
order.
\end{example}

\subsection{Computing winnow}
Various algorithms for evaluating winnow have been proposed in the literature. 
We discuss here those that have a good {\em blocking} behavior and
thus are capable of efficiently processing very large data sets.

We first review BNL (Figure \ref{fig:BNL}), a basic algorithm for evaluating winnow, and then show
that for preference relations that are weak orders a much simpler and
more efficient algorithm is possible.
BNL was proposed in \cite{BoKoSt01} in the context
of {\em skyline queries}. However, \cite{BoKoSt01} also noted
that the algorithm requires only the properties of strict partial orders.
BNL uses  a fixed amount of main memory (a {\em window}).
It also needs a temporary table for the tuples whose status cannot be
determined in the current pass, because the available amount of main memory is limited.
\begin{figure}[htb]
\centering
\fbox{%
\begin{minipage}{.8\textwidth}
\begin{small}\begin{enumerate}
\item clear the window $W$ and the temporary table $F$;
\item make $r$ the input;
\item repeat the following until the input is empty:
\begin{enumerate}
\item for every tuple $t$ in the input:
\begin{itemize}
\item $t$ is dominated by a tuple in $W$ $\Rightarrow$ ignore $t$,
\item $t$ dominates some tuples in $W$ $\Rightarrow$ eliminate the dominated tuples
and insert $t$ into $W$,
\item if $t$ and all tuples in $W$ are mutually indifferent $\Rightarrow$ insert $t$ into $W$
(if there is room), otherwise add $t$ to $F$;
\end{itemize}
\item output the tuples from $W$ that were added there when $F$ was empty,
\item make $F$ the input, clear the temporary table.
\end{enumerate}
\end{enumerate}
\caption{BNL: Blocked Nested Loops}
\label{fig:BNL}
\end{small}\end{minipage}}
\end{figure}

BNL keeps in the window the best tuples discovered so far (some of
such tuples may also be in the temporary table). All the tuples in the
window are mutually indifferent and they all need to be kept,
since each may turn out to dominate some input tuple arriving
later. For weak orders, however, if a tuple $t_1$ dominates $t_2$,
then any tuple indifferent to $t_1$ will also dominate $t_2$. In
this case, indifference is an equivalence relation, and thus it is
enough to keep in main memory only a single tuple $top$ from the
top equivalence class. In addition, one has to keep track of all
members of that class (called the {\em current bucket} $B$), since
they may have to be returned as the result of the winnow. 
Those ideas are behind a new algorithm WWO (Winnow for Weak Orders), shown in Figure
\ref{fig:WWO}.

\begin{figure}[htb]
\centering
\fbox{%
\begin{minipage}{.8\textwidth}
\begin{small}\begin{enumerate}
\item $top$ := the first input tuple
\item $B:=\{top\}$
\item for every subsequent tuple $t$ in the input:
\begin{itemize}
\item $t$ is dominated by $top$ $\Rightarrow$ ignore $t$,
\item $t$ dominates $top$ $\Rightarrow$ $top:=t$; $B:=\{t\}$
\item $t$ and $top$ are indifferent $\Rightarrow$ $B:=B\cup\{t\}$
\end{itemize}
\item output $B$
\end{enumerate}
\caption{WWO: Winnow for Weak Orders} \label{fig:WWO}
\end{small}\end{minipage}}
\end{figure}

It is clear that WWO requires only a single pass over the input.
It uses additional memory (whose size is at most equal to the size
of the input) to keep track of the current bucket. However, this
memory is only written and read once, the latter at the end of the
execution of the algorithm. Clearly, for weak orders WWO is
considerably more efficient than BNL. Note that for weak orders
BNL does not simply reduce to WWO: BNL keeps the mutually indifferent
tuples from the currently top layer in the main memory window (or in the temporary table)
and compares all of them with the input tuple.
The latter is clearly superfluous for preference relations that are weak orders. 
Note also that if additional
memory is not available, WWO can execute in a small, fixed amount
of memory by using two passes over the input: in the first, a top
tuple is identified, and in the second, all the tuples indifferent
to it are selected.

In \cite{ChGoGrLi03} we
proposed SFS, a more efficient variant of BNL for skyline queries,
in which a presorting step is used. Because
sorting may require
more than one pass over the input, that approach will also be less
efficient than WWO for weak orders (unless the input is already sorted).

Even if a preference relation $\succ_C$ is not a weak order in
general, it may be a weak order {\em relative to a class of integrity
constraints $F$}. In those cases, WWO is still applicable.
Note that in such a  case the original definition of $\succ_C$ can still be
used for tuple comparison.

\begin{example}\label{ex:book:05}
Consider Example \ref{ex:book:02}, this time with the $0$-ary FD
$\emptyset \Rightarrow ISBN$. (Such a dependency might hold, for
example, in a relation resulting from the selection $\sigma_{ISBN=c}$
for some constant $c$.)
We already know that the preference relation $\succ_{C_1}$ is a strict partial order.
Being a weak order relative to this FD is captured by
the following CGD:
\[\begin{array}{l}
\forall i_1,v_1,p_1,i_2,v_2,p_2,i_3,v_3,p_3.\\
Book(i_1,v_1,p_1)\wedge Book(i_2,v_2,p_2)\wedge Book(i_3,v_3,p_3)\wedge \phi_1
\Rightarrow \phi_2
\end{array}\]
where 
\[\phi_1:\ (i_1\not= i_2\vee p_1\geq p_2)\wedge (i_2\not= i_3\vee p_2\geq p_3)\]
and 
\[\phi_2:\ (i_1\not= i_3\vee p_1\geq p_3).\]

We show now that this CGD is entailed by the FD $\emptyset \Rightarrow ISBN$.
Assume this is not the case. Then there is an instance of the relation {\it Book}
consisting of tuples $(i_1,v_1,p_1)$,
$(i_2,v_2,p_2)$, and $(i_3,v_3,p_3)$  such 
$\phi_1$ is satisfied but $\phi_2$ is not.
This instance also satisfies the FD, thus $i_1=i_2=i_3$.
We consider the formula $\phi_1\wedge\neg\phi_2$ which can be simplified to
\[p_1\geq p_2\wedge p_2\geq p_3 \wedge p_1< p_3.\]
The last formula is unsatisfiable.
\end{example}

\subsection{Collapsing winnow}

We show here that for weak orders consecutive applications of winnow can be collapsed
to a single one, using prioritized composition. In contrast with Theorem \ref{th:nested},
here we do not impose any conditions on the relationship between the preference relations
involved. Recall that
\[d_1^C:\ \forall t_1,t_2.\ R(t_1)\wedge R(t_2)\Rightarrow t_1\not\succ_{C} t_2.\]

\begin{theorem}\label{th:collapse}
Assume $F$ is a set of integrity constraints over a schema $R$.
If $\succ_{C_1}$ and $\succ_{C_2}$ are preference relations over $R$ such that 
$\succ_{C_1}$ is a weak order relative to $F$,
then for all instances $r\in \sat{F}$:
\[\wnnw{C_1\rhd C_2}(r)=\wnnw{C_2}(\wnnw{C_1}(r)).\]
Additionally, if $\succ_{C_2}$ is a weak order relative to $F\cup d_1^{C_1}$,
then also $\succ_{C_1\rhd C_2}$ is a weak order relative to $F$.
\end{theorem}
\begin{proof}
Let $r\in\sat{F}$.
Assume $t\in\wnnw{C_2}(\wnnw{C_1}(r))$ and $t\not\in\wnnw{C_1\rhd C_2}(r)$.
Then there exists $s\in r$ such that $s\succ_{C_1\rhd\succ_{C_2}} t$.
If $s\succ_{C_1} t$, then $t\not\in\wnnw{C_1}(r)$ and $t\not\in\wnnw{C_2}(\wnnw{C_1}(r))$.
Otherwise, $s\sim_{C_1} t$ and $s\succ_{C_2} t$. If $s\in\wnnw{C_1}(r)$, then 
$t\not\in\wnnw{C_2}(\wnnw{C_1}(r))$. If $s\not\in\wnnw{C_1}(r)$, then
for some $s'\in r$, $s'\succ_{C_1} s$. But then $s'\succ_{C_1} t$ because $\succ_{C_1}$
is a weak order, and consequently $t\not\in\wnnw{C_1}(r)$.

In the other direction, assume $t\in\wnnw{C_1\rhd C_2}(r)$ and $t\not\in\wnnw{C_2}(\wnnw{C_1}(r))$.
If $t\not\in\wnnw{C_1}(r)$, then for some $s\in r$, $s\succ_{C_1} t$.
Thus, $s \succ_{C_1\rhd C_2} t$ and $t\not\in\wnnw{C_1\rhd C_2}(r)$.
If $t\in\wnnw{C_1}(r)$, then for some $s\in \wnnw{C_1}(r)$, $s\sim_{C_1} t$ and $s\succ_{C_2} t$.
Thus again, $s \succ_{C_1\rhd C_2} t$.

The second part of the theorem can be proved in the same way as Proposition 4.6 in \cite{ChTODS03}.
We can require that $\succ_{C_2}$ be a weak order relative to $F\cup d_1^{C_1}$, not just  to $F$,
because the dependency $d_1^{C_1}$ is guaranteed to hold in $\wnnw{C_1}(r)$.
\end{proof}

We show now how Theorem \ref{th:collapse} can be used in query optimization.
Consider the choice between WWO and BNL in the context of Theorem \ref{th:collapse}.
If both $\succ_{C_1}$ and $\succ_{C_2}$ are weak orders (relative to $F$), then it is better
to evaluate $\wnnw{C_1\rhd C_2}(r)$ than $\wnnw{C_2}(\wnnw{C_1}(r))$ because the former
does not require creating intermediate results. In both cases WWO can be used.
If $\succ_{C_2}$ is a strict partial order but not necessarily a weak order (relative to F),
then in both cases we will have to use BNL ($C_1\rhd C_2$ is a strict partial order \cite{ChFOIKS06}), 
so again $\wnnw{C_1\rhd C_2}(r)$ wins.
However, if $\wnnw{C_1}(r)$ is small, it would be better to use WWO to compute $r_1=\wnnw{C_1}(r)$
and then compute $\wnnw{C_2}(r_1)$ using BNL.

Consider now the presence of views.
If $\wnnw{C_1}(R)$ is a non-materialized view, then the query $\wnnw{C_2}(\wnnw{C_1}(R))$
can be first rewritten as $\wnnw{C_1\rhd C_2}(R)$ and then evaluated without the need
for the nested evaluation of $\wnnw{C_1}(R)$.
On the other hand, if $\wnnw{C_1}(R)$ is a materialized view $V$, then it can be used to answer the
query $\wnnw{C_1\rhd C_2}(R)$ by computing  $\wnnw{C_2}(V)$.

\subsection{Further properties}

The list of preference query properties that hold relative to a set of integrity constraints does not end
with those formulated above. There are other algebraic properties that hold conditionally
\cite{ChTODS03}. Such properties can often be formulated in a more general form using CGDs.

For example, consider the commutativity of winnow and selection.
\cite{ChTODS03} shows the following result: 
\begin{proposition}\label{prop:commute}
Given a relation schema $R$, a selection condition $C_1$ over $R$ and a preference formula $C_2$ over $R$,
if the formula 
\[\forall t_1,t_2[(C_1(t_2)\wedge C_2(t_1,t_2))
\Rightarrow C_1(t_1)]\]
is valid, then for all instances $r$ of $R$:
\[\sigma_{C_1}(\wnnw{C_2}(r))=\wnnw{C_2}(\sigma_{C_1}(r)).\]
\end{proposition}

This result can be generalized to hold relative to a set of integrity constraints.
\begin{theorem}\label{th:commute}
Given a relation schema $R$, a selection condition $C_1$ over $R$, a preference formula $C_2$ over $R$,
and a set of integrity constraints $F$ over $R$,
if $\entails{F}{d_2^{C_1,C_2}}$ where

\begin{center}
\fbox{$d_2^{C_1,C_2}:\ \forall t_1,t_2.\ R(t_1)\wedge R(t_2)\wedge C_1(t_2)\wedge C_2(t_1,t_2)\Rightarrow C_1(t_1),$}
\end{center}

then for all instances $r\in\sat{F}$:
\[\sigma_{C_1}(\wnnw{C_2}(r))=\wnnw{C_2}(\sigma_{C_1}(r)).\]
\end{theorem}

\section{Propagation of integrity constraints}\label{sec:propagation}

How do we know whether a specific CGD holds in a relation?
If this is a database relation, then the CGD may be enforced by the DBMS or the application.
If the relation is computed, then we need to determine if the CGD
is preserved in the expression defining the relation.
\cite{Klu80,KlPr82} characterize cases where functional and join
dependencies hold in the results of relational algebra expressions.

We already know that the CGD $d_1^C$ holds in the result of the winnow
$\wnnw{C}$. Winnow  returns a subset of
a given relation, thus it preserves all the CGDs
holding in the relation.
The following theorem characterizes all the dependencies holding in the
result of winnow.

\begin{theorem}\label{th:preserve}
Assume $F$ is a set of CGDs, $f$ a CGD over a schema $R$,
and $\succ_C$ an irreflexive preference relation over $R$.
Then $\entails{F\cup d_1^C}{f}$ iff for every $r\in\sat{F}$,
$\wnnw{C}(r)\in\sat{f}$.
\end{theorem}
\begin{proof}
Assume it is not the case that $\entails{F\cup d_1^C}{f}$.
Thus for some instance $r_0$, $r_0\in\sat{F\cup d_1^C}$
but $r_0\not\in\sat{f}$. Then for all $t_1,t_2\in r_0$, $t_i\sim_C t_j$,
and thus $r_0=\wnnw{C}(r_0)$. Therefore, $\wnnw{C}(r_0)\not\in\sat{f}$.
In the other direction, assume for some $r_0\in\sat{F}$, $\wnnw{C}(r_0)\not\in\sat{f}$.
Thus $r_1=\wnnw{C}(r_0)$ is the instance satisfying $F\cup d_1^C$ and
violating $f$, which provides a counterexample to the entailment of $f$
by $F\cup d_1^C$.
\end{proof}

\begin{example}\label{ex:book:06}
Consider Example \ref{ex:book:02}.
Thus, the FD~ $ISBN\rightarrow Price$ holds in the result of $\wnnw{C_1}$,
because it is entailed by the CGD $d_1^{C_1}$
\[\begin{array}{ll}
d_1^{C_1}:\ &\forall i_1,v_1,p_1,i_2,v_2,p_2,i_3,v_3,p_3.\\
&Book(i_1,v_1,p_1)\wedge Book(i_2,v_2,p_2)\Rightarrow (i_1\not=i_2\vee p_1\geq p_2)
\end{array}\]
even though it might not hold in the input relation {\it Book}.
\end{example}

\section{Computational complexity}\label{sec:complexity}

Here we address the computational issues involved in checking the semantic properties
essential for the semantic optimization of preference queries.
We have shown that such properties can be formulated in terms of the
entailment of CGDs.
We assume that we are dealing with $k$-dependencies
for some fixed $k\geq 1$. For example, for FDs $k=2$.
Notice also that all the interesting properties studied in this paper, e.g., containment or weak order,
can be expressed as $k$-dependencies for $k\leq 3$.

We assume here that the CGDs under consideration are {\em clausal}: the constraint in the
body is a conjunction of atomic constraints and the head consists of a disjunction of  atomic constraints.
All the dependencies that we have found useful in the context of semantic
optimization of preference queries are clausal.

\subsection{Upper bounds}

\cite{BaChWo99} shows a reduction from the entailment of CGDs to 
the validity of universal formulas in the constraint theory.
The basic idea is simple: the entailment of $k$-dependencies needs
to be considered over relation instances of cardinality at most $k$,
and each such instance can be represented by $k$ tuple variables.
Each dependency $f$ is mapped to a constraint formula $cf_k(f)$.
Then the entailment of a CGD $f_0$ by a set of CGDs $F$ is expressed
as the validity of the formula:
\[\forall^*. (\bigwedge_{f\in F}cf_k(f))\Rightarrow cf_k(f_0),\]
or equivalently, as the unsatisfiability of a quantifier-free CNF formula obtained from its negation.

The following example illustrates the construction of $cf_k(f)$.
\begin{example}
Consider the dependency: 
\[d_0^{C_1,C_2}:\ \forall t_1,t_2.\ R(t_1)\wedge R(t_2)\wedge t_1\succ_{C_1} t_2\Rightarrow t_1\succ_{C_2} t_2.\]
We have that $cf_2(d_0^{C_1,C_2})$ is equal to
\[\begin{array}{l}
[C_1(t_1,t_1)\Rightarrow C_2(t_1,t_1)] \wedge [C_1(t_2,t_2)\Rightarrow C_2(t_2,t_2)]\\
\wedge [C_1(t_1,t_2)\Rightarrow C_2(t_1,t_2)] \wedge [C_1(t_2,t_1)\Rightarrow C_2(t_2,t_1)]\\\end{array}\]
which for irreflexive $\succ_{C_1}$ is equivalent to
\[[C_1(t_1,t_2)\Rightarrow C_2(t_1,t_2)] \wedge [C_1(t_2,t_1)\Rightarrow C_2(t_2,t_1)].\]
\end{example}
For a fixed $k$, the size of $cf_k(f)$ is linear in the size of $f$.
Thus we can easily characterize the complexity of dependency entailment.

\begin{theorem}\label{th:expo}
Assume $F$ is a set of $k$-CGDs for a fixed $k\geq 1$ over a constraint
theory of equality/rational-order constraints, and preference relations are defined
by ipfs over the same constraint theory.
Checking containment, dependency propagation, and weak or strict partial order property, relative to  $F$,
are all in co-NP.
\end{theorem}
\begin{proof}
Satisfiability of conjunctions of atomic constraints in this constraint theory can be checked
in polynomial time \cite{GuSuWe96}.
Thus satisfiability of quantifier-free formulas in this constraint theory is in NP.
\end{proof}

What remains now to be shown is that (1) the intractability is, in general, unavoidable;
and (2) special tractable cases exist.

\subsection{Lower bounds}

\cite{BaChWo99} show a number of co-NP-completeness results for the entailment problem restricted
to special classes of CGDs. To adopt those results to the context of the semantic properties of preference queries
studied in the present paper, we need to show that the hardest (co-NP-hard) cases of the entailment
can be equivalently expressed in terms of such properties.
Such an approach is adopted in the proofs of Theorems
\ref{th:redundant:complexity} and \ref{th:preserve:complexity} to characterize the complexity
of testing redundancy of winnow and propagation of integrity constraints.
On the other hand, in Theorem \ref{th:weak:complexity}, a new reduction is introduced for the problem of testing
the (relative) weak order property.

\begin{theorem}\label{th:redundant:complexity}
Checking whether $\wnnw{C}$ is redundant
relative to $F$, where $F$ is a set of $2$-CGDs and $C$ is a rational-order ipf defining a strict partial order, is
co-NP-hard.
\end{theorem}
\begin{proof}
We adapt the proof of Theorem 4.3 in \cite{BaChWo99}.
The reduction there is from SET SPLITTING but the same reduction applies
to MONOTONE 3-SAT. Assume we are given a propositional formula $\phi$ with $n$ variables
$p_1,\ldots,p_n$, consisting
of $l$ positive clauses of the form $c_h\equiv p_i\vee p_j\vee p_m$, $h=1,\ldots,l$, and $k$ negative clauses
$c_h\equiv\neg p_i\vee\neg p_j \vee \neg p_m$, $h=1,\ldots,k$.
We consider a relation $R$ with $n+k+1$ attributes.
The truth of a propositional variable $p_i$ is represented by equality on the attribute $i$.
We build the set $F$ of $2$-CGDs in stages.
A positive clause $p_i\vee p_j\vee p_m$ is mapped to a CGD
\[\forall t_1,t_2.\ R(t_1)\wedge R(t_2)\wedge t_1[i]\not=t_2[i]\wedge t_1[j]\not=t_2[j]
\Rightarrow t_1[m]=t_2[m].\]
The construction for a  negative clause $c_h\equiv\neg p_i\vee\neg p_j \vee \neg p_m$ is more complicated.
We construct the following FDs:
\[\begin{array}{l}
\forall t_1,t_2.\ R(t_1)\wedge R(t_2)\wedge t_1[i]=t_2[i]\wedge t_1[n+h]=t_2[n+h]\\
\hspace{20pt}\Rightarrow t_1[n+k+1]=t_2[n+k+1], \\
\forall t_1,t_2.\ R(t_1)\wedge R(t_2)\wedge t_1[j]=t_2[j]\wedge t_1[m]=t_2[m]
\Rightarrow t_1[n+h]=t_2[n+h].
\end{array}\]
Finally, we define the preference relation $\succ_C$:
\[t_1\succ_C t_2 \equiv t_1[n+k+1]>t_2[n+k+1].\]
Thus the CGD $d_1^C$ is
\[\forall t_1,t_2.\ R(t_1)\wedge R(t_2)\Rightarrow t_1[n+k+1]\leq t_2[n+k+1].\]
Along the same lines as in the proof in \cite{BaChWo99}, we can show
that $\phi$ is unsatisfiable iff $\entails{F}{d_1^C}$.
\end{proof}

\begin{theorem}\label{th:preserve:complexity}
Checking whether $\entails{F\cup d_1^C}{f}$, 
where $F$ is a set of $2$-CGDs and $C$ is a rational-order ipf defining a strict partial order,
is co-NP-hard.
\end{theorem}
\begin{proof}
We modify the proof of Theorem \ref{th:redundant:complexity}.
We pick one  positive clause $p_i\vee p_j\vee p_m$. It is still  mapped to a CGD equivalent to the
previous one:
\[\forall t_1,t_2.\ R(t_1)\wedge R(t_2)\Rightarrow  t_1[i]=t_2[i]\vee t_1[j]=t_2[j]
\vee t_1[m]=t_2[m]\]
but this CGD is now obtained as the special dependency $d_1^C$ for $\succ_C$ defined
as follows
\[t_1\succ_C t_2 \equiv t_1[i]\not=t_2[i]\wedge t_1[j]\not=t_2[j]
\wedge t_1[m]>t_2[m].\]
The CGD $f$ is:
\[f:\ \forall t_1,t_2.\ R(t_1)\wedge R(t_2)\Rightarrow t_1[n+k+1]=t_2[n+k+1].\]
The construction for the remaining positive clauses, as well as all the negative clauses, remains the same.
\end{proof}

\begin{theorem}\label{th:weak:complexity}
Checking whether $\succ_C$ is a weak order
relative to  $F$, where $F$ is a set of $3$-CGDs and $C$ is an equality ipf defining a strict partial order, is
co-NP-hard.
\end{theorem}
\begin{proof}
Reduction from 3-colorability.
Assume we are given a graph $G=(V,E)$ where $V=\{v_1,\ldots,v_n\}$.
We construct the set $F$ consisting of the following CGDs:
\[\begin{array}{l}
\forall t.\ R(t)\Rightarrow t[i]=0 \vee t[i]=1,\\
\forall t.\ R(t)\Rightarrow t[n+1]=1 \vee t[n+1]=2\vee t[n+1]=3,\\
\forall t_1,t_2,t_3.\ R(t_1)\wedge R(t_2)\wedge R(t_3)\wedge
t_1[n+1]\not= t_2[n+1]\\\hspace{20pt}\wedge t_1[n+1]\not= t_3[n+1]\wedge t_2[n+1]\not= t_3[n+1]
\Rightarrow \gamma(t_1[i],t_2[i],t_3[i]),\\
\end{array}\]
where $i=1,\ldots n$ and $\gamma(x,y,z)$ is a formula saying that exactly one of $x$, $y$ and $z$ is equal to $1$.
The last dependency is not clausal but can easily be represented as a set of clausal CGDs.
Also, for every edge $(v_i,v_j)\in E$, we include the following CGD:
\[\begin{array}{l}\forall t_1,t_2,t_3.\ R(t_1)\wedge R(t_2)\wedge R(t_3)\wedge
t_1[n+1]\not= t_2[n+1]\wedge t_1[n+1]\not= t_3[n+1]\\\hspace{20pt}\wedge t_2[n+1]\not= t_3[n+1]
\Rightarrow t_1[i]\not=t_1[j]\vee t_2[i]\not= t_2[j]\vee t_3[i]\not= t_3[j].
\end{array}\]
Finally, we define the strict partial order $\succ_C$ as follows:
\[t\succ_C t'\equiv t[n+1]=1\wedge t'[n+1]=2.\]

Assume now that $G$ is $3$-colorable. We construct an instance $r=\{t_1,t_2,t_3\}$ as follows.
We will use $t_k$, $k=1,\ldots, 3$, to represent the vertices colored with the color $k$.
We make $t_k[i]=1$ if $v_i$ is colored with $k$; $t_k[i]=0$ otherwise.
We make $t_1[n+1]=1$, $t_2[n+1]=2$ and $t_3[n+1]=3$.
By construction, $r$ satisfies $F$ but $\succ_C$ on $r$ is not a weak order.
Therefore, $\succ_C$ is not a weak order relative to $F$.

In the other direction, take an instance $r=\{t_1,t_2,t_3\}$ satisfying $F$ but such that
$\succ_C$ on $r$ is not a weak order. Then $\{t_1[n+1],t_2[n+1],t_3[n+1]\}=\{1,2,3\}$.
Then $r$ encodes a $3$-coloring for $G$.
 \end{proof}

\subsection{Tractable cases}

We obtain our first tractability results by identifying a new  case of PTIME entailment.
The case involves the entailment of a CGD over equality constraints by a set of FDs.
This case was not studied in \cite{BaChWo99}.
Note that it is more general than the standard FD entailment because the CGD may contain
general equality constraints.

\begin{theorem}\label{th:equality}
Let $F=\{f_1,\ldots,f_n\}$ be a set of FDs and $f_0$ a clausal $k$-CGD over equality constraints.
Then  checking  whether $F \models f_0$ is in PTIME.
\end{theorem}
\begin{proof}
The dependency $f_0$ is of the form 
\[\forall t_1.\ldots\forall t_k.\ [R(t_1)\wedge\cdots\wedge R(t_k)\wedge \gamma(t_1,\ldots t_k)]
\Rightarrow \gamma'(t_1,\ldots t_k).\]
As explained earlier in this section,
the entailment $F \models f$ reduces to the validity of the formula
\[\forall t_1,\ldots,t_k. (\bigwedge_{f\in F}cf_k(f))\Rightarrow cf_k(f_0),\]
which is the same as the unsatisfiability of the formula
\[(\bigwedge_{f\in F}cf_k(f))\wedge\neg cf_k(f_0).\]
We note that for any fd $f\equiv X\rightarrow Y$, $cf_k(f)$ is a conjunction $E$
of implications
\[\bigwedge_{i,j=1,\ldots,k}t_i[X]=t_j[X]\Rightarrow t_i[Y]=t_j[Y].\]

On the other hand, $\neg cf_k(f_0)$ is a disjunction of conjunctions $S_1,\ldots, S_m$ of atomic equality
constraints.
Each $S_1$, $i=1,\ldots, m$, is of the form $\phi(t_{i_1},\ldots, t_{i_k})\wedge
\psi(t_{i_1},\ldots, t_{i_k})$ where $i_1,\ldots,i_k\in\{1,\ldots,k\}$,
$\phi(t_{i_1},\ldots, t_{i_k})$ is a conjunction of equalities,
and $\psi(t_{i_1},\ldots, t_{i_k})$ is a conjunction of inequalities.
Both of those conjunctions can be viewed as sets of atomic constraints.

To determine the satisfiability of the formula $E\wedge (S_1\vee\cdots\vee S_m)$,
we need to check whether $E\wedge S_i$ is satisfiable for any $i=1,\ldots,m$.
This can be done by essentially propositional Horn reasoning.
We encode each equality and inequality by a different propositional variable
and add Horn clauses representing the transitivity, symmetry and reflexivity of equality.
Using those clauses together with the implications in $cf_k(f)$ for $f\in F$,
we then derive all the equalities implied by those in $S_i$
and check whether any of them violates reflexivity or conflicts with an inequality in $S_i$.
The satisfiability of $E\wedge S_i$ can thus be determined in polynomial time.
\end{proof}

\begin{corollary}
Given a set of FDs F and equality ipfs $C_1$ (in DNF) and $C_2$ (in CNF), the
following properties can be checked in PTIME:
\begin{enumerate}
\item the containment of $\succ_{C_1}$ in $\succ_{C_2}$ relative to $F$, and
\item $\succ_{C_1}$ being a weak or strict partial order relative to $F$.
\end{enumerate}
\end{corollary}

The requirement that the formulas $C_1$ and $C_2$ be in an appropriate normal form
guarantees that the dependencies $d_0^{C_1,C_2}$ and $d_1^{C_1}$ are representable
using polynomially many clausal CGDs.

We obtain here further tractable cases of the semantic properties studied in the present paper
by adapting the results of 
\cite{BaChWo99}. That paper identifies several classes of CGDs for which the entailment problem is tractable.

The restrictions we impose on CGDs and preference formulas may be of the following kinds:
\begin{itemize}
\item the atomic constraints should be {\em typed};
\item the number of atomic constraints should be bounded;
\item the {\em width} and the {\em span} of preference formulas, defined below, should be restricted.
\end{itemize}
 
\begin{definition}
A constraint formula $C(t_1,\ldots,t_n)$ over tuple variables $t_1,\ldots,t_n$ 
is {\em typed} if all its atomic subformulas are of the form $t_i[A]\theta t_j[A]$
or $t_i[A]\theta c$, where $A$ is an attribute, $c$ a constant, and $\theta\in\{=,\not=,<,>,\leq,\geq\}$.
A CGD is typed if all its constraints are typed.
\end{definition}

The size of a preference
formula $C$ (over a relation $R$) in DNF
is characterized by two parameters: ${\it width\/}(C)$ -- the number of disjuncts in $C$,
and ${\it span\/}(C)$ -- the maximum number of conjuncts in a disjunct of $C$.
Namely, if $C=D_1\vee\cdots\vee D_m$, and each $D_i=C_{i,1}\wedge\cdots C_{i,k_i}$,
then ${\it width\/}(C)=m$ and ${\it span\/}(C)=\max \{k_1,\ldots,k_m\}$.

Consider first the containment problem. To check whether $\contains{F}{\succ_{C_1}}{\succ_{C_2}}$, we need to determine
whether $\entails{F}{d_0^{C_1,C_2}}$ where
\[d_0^{C_1,C_2}:\ \forall t_1,t_2.\ R(t_1)\wedge R(t_2)\wedge t_1\succ_{C_1} t_2\Rightarrow t_1\succ_{C_2} t_2.\]
To obtain tractability we need to impose simultaneous restrictions on $F$, $\succ_{C_1}$, and $\succ_{C_2}$.

\begin{theorem}
Let $F$ be a set of typed clausal $2$-CGDs with two atomic constraints over a schema $R$, and $C_1$ and $C_2$ typed 
preference formulas
over the same schema and the same constraint theory (either equality or rational order).
Moreover, none of $F$, $C_1$, and $C_2$ contains constants.
Then 
\begin{itemize}
\item checking whether $\ \contains{F}{\succ_{C_1}}{\succ_{C_2}}\ $ can be done in PTIME if ${\it span\/}(C_1)\leq 1$ and ${\it width\/}(C_2)\leq 1$, and 
\item checking whether $\ \contains{F}{\succ_{C_1}}{\succ_{False}}\ $ can be done in PTIME if ${\it span\/}(C_1)\leq 2$.
\end{itemize}
\end{theorem}

Note that, for example, unary FDs are typed $2$-CGDs with two atomic equality constraints.

Consider now the problem of propagating integrity constraints.

\begin{theorem}
Let $F$ be a set of clausal $k$-CGDs, $f$ a clausal $k$-CGD and $C$ a preference formula over the same schema,
and none of $F$, $f$, and $C$ contains constants.
Then checking whether $\entails{F\cup d_1^C}{f}$ can be done in PTIME if:
\begin{itemize}
\item $F$, $f$, and $\neg C$ have at most one atomic constraint each, or
\item $k=2$, and $F$, $f$ and $\neg C$ are typed and contain each at most two atomic
constraints over the same constraint theory (either equality or rational order).
\end{itemize}
\end{theorem}

The results of \cite{BaChWo99} cannot be applied to identify tractable cases of the weak order
or the strict partial order property, because those properties are formulated using $3$-CGDs
with three or more atomic constraints. Such CGDs do not fall into any of the tractable classes
of \cite{BaChWo99}.

\section{Related work}\label{sec:related}
The basic reference for semantic query optimization is
\cite{CHGrMi90}. The most common techniques are: join
elimination/introduction, predicate elimination and introduction, and
detecting an empty answer set. \cite{GryzVLDB99} discusses the
implementation of predicate introduction and join elimination in
an industrial query optimizer. Semantic query optimization
techniques for relational queries are studied in \cite{ZhOz97}
in the context of denial and referential constraints,
and in \cite{MaWa00} in the context of constraint
tuple-generating dependencies (a generalization of CGDs and classical
relational dependencies).
FDs are used for reasoning about
sort orders in \cite{SiShMa96}.

Two different approaches to preference queries have been pursued
in the literature: qualitative and quantitative. In the {\em
qualitative} approach, represented by
\cite{LaLa87,KiGu94,KoKiThGu95,BoKoSt01,GoJaMa01,ChEDBT02,ChTODS03,Kie02,KiHa02,KiKo02},
the preferences between tuples in the answer to a query are
specified directly, typically using binary {\em preference
relations}. In the  {\em quantitative} approach, as represented by
\cite{AgWi00,HrPa04}, preferences are specified indirectly using
{\em scoring functions} that associate a numeric score with every
tuple of the query answer. Then a tuple $t_1$ is preferred to a
tuple $t_2$ iff the score of $t_1$ is higher than the score of
$t_2$. The qualitative approach is strictly more general than the
quantitative one, since one can define preference relations in
terms of scoring functions However, not every intuitively
plausible preference relation can be captured by scoring
functions.
\begin{example}\label{ex:book:01}
There is no scoring
function that captures the preference relation described in Example \ref{ex:book}.
Since there is no preference defined between any of the first three tuples
and the fourth one, the score of the fourth tuple should be equal to all of
the scores of the first three tuples. But this implies that the scores
of the first three tuples are the same, which is not possible since
the second tuple is preferred to the first one which in turn is preferred
to the third one.
\end{example}
\begin{example}
Another common example of a preference relation that is not representable using
a utility function is the {\em threshold of detectable difference} relation $\succ_t$:
\[x\succ_t y \equiv x\geq y+c\]
where $c$ is the threshold value ($c>0$).
\end{example}
This lack of expressiveness of the quantitative approach is well known
in utility theory \cite{Fish99,Fish70}.
The importance of weak orders in this context comes from the fact that only weak orders
can be represented using real-valued scoring functions (and for countable domains
this is also a sufficient condition for the existence of such a representation \cite{Fish70}).
However, even if a utility function is known to exist, its definition may be non-explicit
\cite{Fish70} and thus unusable in the context of database queries.
In the present paper we do not assume that preference relations are weak orders.
We only characterize a condition under which preference relations become weak orders
relative to a set of integrity constraints.
In such cases, we can exploit the benefits of a preference relation being a (relative) weak order,
for example the possibility of using WWO for computing winnow, without a need to construct
a specific utility function representing the preference relation.

Algebraic optimization of preference queries is discussed in the papers \cite{ChTODS03,KiHa02,KiHa03}.

\section{Conclusions and further work}\label{sec:concl}
We have presented several novel techniques for semantic optimization of
preference queries, focusing on the winnow operator.
We characterized the necessary and sufficient conditions
for the applications of those techniques in terms of the
entailment of constraint-generating dependencies.
(This idea was suggested but not fully developed in \cite{ChCDB04}.)
As a consequence, we were able to leverage some of the computational complexity
results from \cite{BaChWo99}.
Moreover, we proved here several new complexity results:
Theorems \ref{th:weak:complexity} and \ref{th:equality}.
Theorem \ref{th:collapse} is also completely new.
Other results are reformulations of those presented in \cite{ChCDB04}.

The simplicity of our results attests to the power of logical
formulation of preference relations. However, our results are
applicable not only to the original logical framework of
\cite{ChEDBT02,ChTODS03}, but also to preference queries defined
using preference constructors \cite{Kie02,KiKo02} and skyline
queries \cite{BoKoSt01,ChGoGrLi03,KoRaRo02,PaTaFuSe03} because
such queries can be expressed using preference formulas.

\ignore{
For the main results of this paper to hold for a class of integrity constraints, two
conditions need to be satisfied: (a) the class should be able to
express CGDs expressing the semantic properties of interest;
and (b) the notions of entailment and finite entailment (entailment
on finite relations) for the
class should coincide. If (b) is not satisfied,
then the results will still hold if  reformulated by replacing
"entailment" with "finite entailment". Thus, assuming that
(a) is satisfied, the effectiveness of checking the preconditions
of the above theorems depends on the decidability of finite
entailment for the given class of integrity constraints.
}

The ideas presented in this paper in the context of winnow can be adapted
to other preference-related operators.
For example, {\em ranking} \cite{ChTODS03} associates with each tuple
in a relation its rank. The best tuples (computed by winnow) have rank 1,
the second-best tuples have rank 2, etc.
The algorithm WWO can be extended to compute ranking instead of winnow,
and thus the computation of ranking will also benefit if the given preference relation is a
weak order relative to the given integrity constraints.

Further work can address, for example, the following issues:
\begin{itemize}
\item identifying other semantic optimization
techniques for preference queries,
\item expanding the class of integrity constraints by considering, e.g., tuple-generating
dependencies and referential integrity constraints,
\item deriving further tractable cases of (relative) containment and satisfaction of order axioms,
\item studying the preservation of general constraint-generating dependencies by relation
algebra operators and expressions (\cite{KlPr82} consider this problem for functional and join dependencies);
\item identifying weaker but easier to check sufficient conditions for the application
of our techniques.
\end{itemize}
\bibliographystyle{plain}

\end{document}